\documentclass{optica-article}
\journal{opticajournal} 
\articletype{Research Article}
\usepackage{mathbbol}
\usepackage{float}
\usepackage{physics}
\usepackage{graphicx}
\usepackage{dcolumn}
\usepackage[normalem]{ulem}
\usepackage{bm}
\usepackage{subcaption}
\usepackage{booktabs}
\usepackage{cleveref}
\usepackage{soul}

\begin{document}

\title{Loss-resilient, efficient x-ray interaction-free measurements}

\author{Ron Cohen,\authormark{1} Sharon Shwartz,\authormark{2} and Eliahu Cohen\authormark{1,*}}

\address{
\authormark{1}
Faculty of Engineering and Institute of Nanotechnology and Advanced Materials, Bar-Ilan University, Ramat Gan 5290002, Israel
\\
\authormark{2}
Department of Physics and Institute of Nanotechnology and Advanced Materials, Bar-Ilan University, Ramat Gan 5290002, Israel
}

\email{\authormark{*}eliahu.cohen@biu.ac.il} 

\begin{abstract*} 
Interaction-free measurement (IFM) is a promising technique for low-dose detection and imaging, offering the unique advantage of probing an object without absorption of the interrogating photons. 
We propose an experiment to demonstrate IFM  
in the single x-ray photon regime.
The proposed scheme relies on the triple-Laue (LLL) symmetric x-ray interferometer, where each Laue 
diffraction 
acts as a lossy beamsplitter. 
In contrast to many quantum effects which are highly vulnerable to loss, we show that an experimental demonstration of this effect in the x-ray regime is feasible and can achieve high IFM efficiency even in the presence of substantial loss in the system. The latter aspect is claimed to be a general property of IFM based on our theoretical analysis.
We scrutinize
two suitable detection schemes that offer efficiencies of up to $\eta\sim \frac{1}{2}$. The successful demonstration of IFM with x-rays promises intriguing possibilities for measurements with reduced dose, mainly advantageous for biological samples, where radiation damage is a significant limitation.
\end{abstract*}

\section{Introduction}
In quantum mechanics, the collapse or reduction of the wave function due to measurement may occur even in cases of ``nondetection'' events, where there appears to be no interaction between the measurement device and the system. The origin of this idea is rooted in Renninger's ``negative-result'' experiment \cite{renninger1960messungen}.
The notion of an ``interaction-free'' quantum measurement, first coined by Dicke \cite{dicke1981interaction}, describe a scenario in which a non-scattering event of a photon alters the wave function of an atom, although the quantum state of the electromagnetic field has not been affected. 
Later works introduced interaction-free measurement (IFM) \cite{elitzur1993quantum,vaidman1994realization,kwiat1995interaction} in a somewhat different and more applicable context, for which detection of an object is possible, without absorption or scattering of the probing particle by the object.
This uniquely quantum effect, known also as the Elitzur-Vaidman bomb tester, relies on the distinction between a scenario in which a particle interferes with itself and when it takes a certain path and thus does not interfere. Therefore, the presence of an object which obstructs the interference can be inferred by the detectors and thus an object can be detected without interacting with the particle.
Initial experimental demonstrations of this effect \cite{kwiat1995interaction} achieved IFM efficiencies of $\eta<\frac{1}{2}$ and later demonstrations \cite{kwiat1998experimental, tsegaye1998efficient} achieved high IFM efficiencies ($\eta>\frac{1}{2}$), by utilizing the quantum Zeno effect
(i.e., 
repeated interrogation \cite{kwiat1995interaction}).
These demonstrations inspired the development of IFM-based low-dose imaging techniques \cite{white1998interaction,zhang2019interaction,pualici2022interaction}. 
Given the delicate nature of biological samples, the utilization of IFM, particularly x-ray IFM, presents a compelling opportunity.

There are several advantages to employing IFM with x-rays for biomedical imaging.
The non-invasive capabilities of x-rays make them a  unique tool for imaging
of internal biological structures, including bones and organs.
Furthermore, commercially available x-ray detectors can reach nearly 100\% efficiency with low dark current and photon number resolving capabilities over a very broad spectral range \cite{scuffham2012cdte,send2013characterization}.
The primary method for generating single x-ray photons is through heralding using spontaneous parametric down-conversion (SPDC) \cite{sofer2019quantum, strizhevsky2021efficient}.
Radioactive sources of M\"ossbauer nuclei with a cascade scheme provide another method for generating and controlling single $\gamma$-ray photons
\cite{vagizov2014coherent}.
In addition, quantum recoil was recently suggested as a promising method for generating tunable single x-ray photons \cite{huang2023quantum}.
Furthermore, attenuated x-ray coherent source can also be used to generate single photons by post-selection.
Also, there 
is a wide variety of
x-ray interferometers available \cite{hart1975review, lider2014x, liss2000storage, chang2005x}.
For the purpose of x-ray IFM, we note 
the triple-Laue (LLL) \cite{bonse1965x} and Fabry-P\'{e}rot (FB) \cite{liss2000storage, chang2005x}. The former
can be used
for a proof-of-concept demonstration as discussed below, and the latter can be employed for implementing high-efficiency IFM \cite{tsegaye1998efficient,namekata2006high}.
We emphasize that at first glance, there appears to be a potentially significant challenge for the implementation of x-ray IFM. Absorption has proven to be significant in almost all previously reported x-ray interferometers, 
typically 
reducing the efficiency of quantum effects.
However, we show that IFM exhibits notable resilience to losses, unlike many other quantum effects and technologies. Therefore, the utilization of x-ray IFM holds particular promise in this regard.
A first experimental demonstration of x-ray IFM can pave the way for the development of low-dose bio-imaging schemes.\\
Here we analyze the feasibility of demonstrating this effect in a lossy LLL system, in the single x-ray photon regime.
We theoretically modeled the LLL system as being comprised of 4 identical lossy beamsplitters, such that each Laue diffraction acts as a beamsplitter.
We identified two possible detection schemes, characterized their efficiencies and concluded that an x-ray IFM demonstration is feasible even in the lossy regime.

\section{Proposed experimental scheme}
\begin{figure*}[h!]
	\begin{center}
	\subfloat[
	\centering 
	Triple-Laue symmetric interferometer
	]{%
		\includegraphics[scale=0.52]{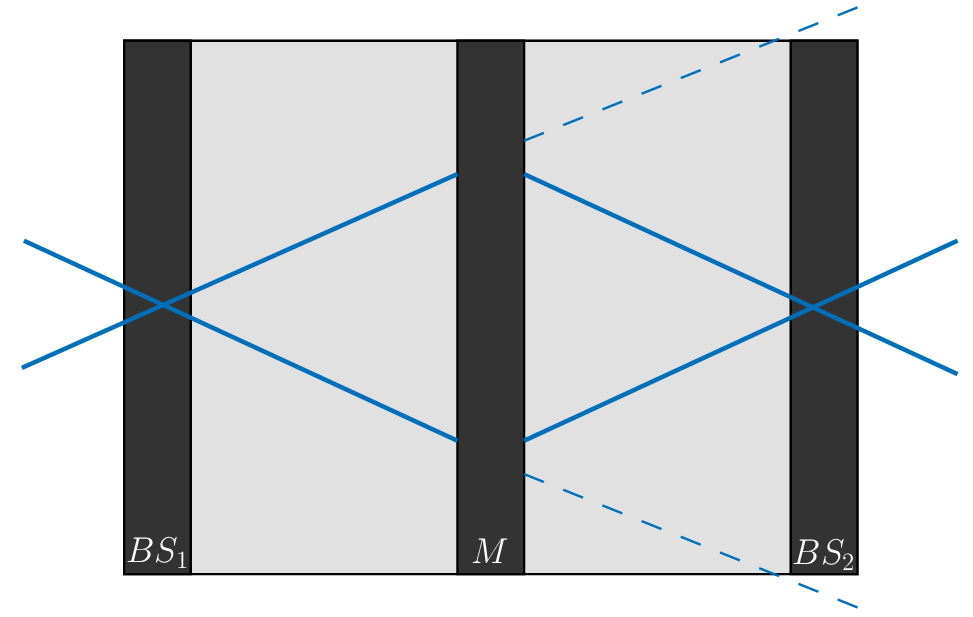}%
	}
	\hfill
	\subfloat[
	\centering 
	Theoretical representation of the LLL system
	]{%
		\includegraphics[scale=0.6]{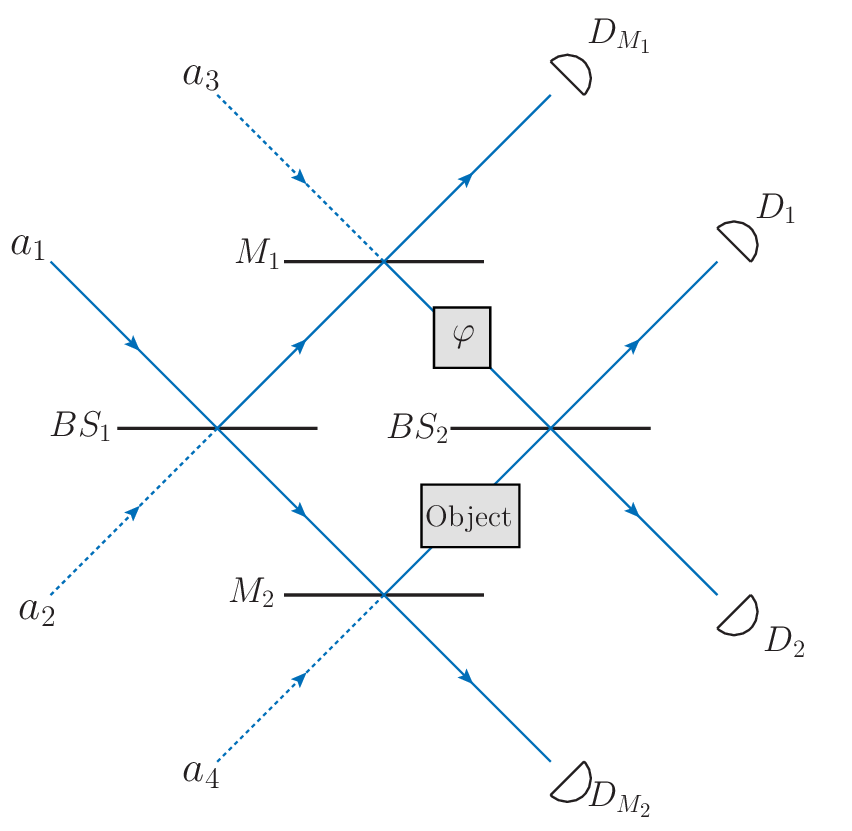}%
	}
	\caption{
		\textbf{(a)} LLL system: The first plate ($BS_1$) acts as the first beamsplitter. The second plate ($M$) acts as two mirrors.
		The third plate ($BS_2$) acts as the second beamsplitter.
		\textbf{(b)} Theoretical representation of the LLL system:
		Input operators undergo first beamsplitter ($BS_1$).
		Each beam is then reflected by the two mirrors ($M_1,\,M_2$) and each such mirror can be considered to be a beamsplitter.
		The reflected beam from the first mirror undergoes a phase shift $\varphi$, and 
		finally passes through the second beamsplitter ($BS_2$).
		Thereafter,
		the reflected beam from the second mirror is completely absorbed by an object.
		The dashed lines indicate vacuum input.		
	}%
	\label{fig:LLLsystem}%
\end{center}
\end{figure*}

Our proposal is based on the LLL interferometer \cite{bonse1965x}, which 
operates similarly 
to the Mach-Zehnder interferometer but specifically designed for x-rays. 
The LLL interferometer is constructed by cutting two wide grooves in a single crystal block to form three plates, as depicted in Fig. \ref{fig:LLLsystem} (a).
The first plate operates as a beamsplitter, the second plate operates as two mirrors and the third plate is another beamsplitter.
In each plate a Laue diffraction occurs, such that different portions of the beam are transmitted, diffracted and absorbed.
The ratio between the transmission and diffraction is determined by the width of the plates, incident angle and the spectral profile of the beam.
Therefore, control of the ratio of transmission and reflection is possible, by tuning the incident angle with respect to Bragg's angle.
Controlling this ratio is crucial for optimizing IFM efficiency, a point that will be further elucidated in the next section.	
The same system can be theoretically described  as a Mach-Zehnder interferometer, in which each mirror is replaced by a beamsplitter,
as depicted in Fig. \ref{fig:LLLsystem} (b).
The system consists of four identical beamsplitters ($BS_1, BS_2,M_1,M_2$), with
identical transmission ($\tilde{T}$), reflection ($\tilde{R}$) and 
loss
($\tau$)
coefficients.
We define the lossless transmission and reflection coefficients as ($T, R$) respectively, such that 
\begin{equation}\label{RT-coeff}
	\begin{aligned}
		&
		\tilde{R}=\tau R, && \tilde{T}=\tau T,
		\\&
		R+T=1, && \tilde{R}+\tilde{T}=\tau,
	\end{aligned}
\end{equation}
where the parameter $\tau$ determines the loss, such that it is lossless for $\tau=1$ and otherwise $0\leq\tau<1$.
In addition, we consider a possible phase term ($P(\varphi)$) and the possibility of an object lying on one of the arms.
\\
We utilize the Heisenberg picture, in which we evolve in time the input operators.
In our scheme, we consider 4 input operators.
The two input operators for $BS_1$ are $a_1$ and $a_2$, and the operators for the vacuum inputs of the mirrors are $a_3$ ($M_1$) and $a_4$ ($M_2$).
Except for the phase term $P(\varphi)$, all other optical components are lossy and therefore, introduce noise.
Consequently, noise operators are included in the transformation of each lossy component, and their magnitude is proportional to the loss.
Theoretically, each optical component is described by a linear transformation acting on an input operator vector $\vb*{a}_{in}$.
\begin{equation}\label{key}
	\begin{aligned}
		&
		\begin{aligned}
			\vb*{a}_{in}
			=
			\mqty(
			\vb*{a}
			\\
			\vb*{a}_{n})
			,
		\end{aligned}
		\\&
		\begin{aligned}
			&
			\vb*{a}
			=\mqty(a_3
			\\
			a_1
			\\
			a_2
			\\
			a_4
			),
			&&
			\vb*{a}_{n}
			=
			\mqty(a_{n1}
			\\
			a_{n2}
			\\
			\vdots
			\\
			a_{n8}
			).
		\end{aligned}
	\end{aligned}
\end{equation}
The input operator vector $\vb*{a}_{in}$ includes our physical input and noise operators denoted as $\vb*{a}, \vb*{a}_n$, respectively.
\\
Schematically, each optical component (OC) can therefore be expressed as
\begin{equation}\label{key}
	OC
	=\mqty(\sqrt{\tau}U_{OC} & \sqrt{1-\tau}n_{OC}
	\\ \mathbb{0} & \mathbb{1}
	),
\end{equation}
where $U_{OC}$ represents an optical component acting on the input modes ($a_1, a_2, a_3, a_4$) weighted by the square root of the loss parameter, and
to introduce vacuum noise (for $\tau<1$), we employ $n_{OC}$ to distribute noise operators for the relevant modes (for more details, see Eq. (S14, S15, S16)
in 
the Supplement 1.
Clearly, the added noise is proportional to the loss e.g., for the lossless case ($\tau=1$) the added noise vanishes.
\\
For this system, there are 5 possible transformations, one for each beamsplitter/mirror and for the phase term ($BS_1, BS_2,M_1,M_2,P(\varphi)$); their explicit form is stated in 
the first part of
the Supplement 1.
Realistic representation of the optical components presented here, can be effectively realized by modeling Laue diffraction as a lossy beamsplitter transformation as shown in
Sec.
\ref{section:Laue-beamsplitter}.
\section{Detection configurations}
IFM detection relies on statistical contrast in output port measurements between 
scenarios in which an object is present and absent.
The best possible contrast is achieved when we have an ideal dark port (in the absence of an object), namely, a port for which the detector will never measure a photon due to complete destructive interference.
In this case, a photon can be detected at our designated dark port only if an object obstructs the interference. 
Since the object creates path distinguishability for the photon,  object detection can be achieved without it absorbing any radiation.
To quantify the efficiency of our detection, we 
use the conventional IFM efficiency \cite{kwiat1995interaction} figure of merit 
\begin{equation}\label{IFM-efficiency}
	\eta=
	\frac{P(\text{detection})}{P(\text{detection})+P(\text{absorption})},
\end{equation}
where $P(\text{detection})$ is the probability to detect an object without absorption by the object and $P(\text{absorption})$ is the probability for the object to absorb the photon.
This figure of merit was originally conceived for an ideal (lossless) system, for which it quantifies the portion of detection events that are interaction-free (since absorption indicates the presence of an object). However, it still bears meaning in the general (lossy) case, since we are only concerned with absorption events by the object which increase the radiation dose.
\\\\
Generally, depending on the type of interferometer used, there can be several configurations for which a dark port is achieved, and for each configuration there is a different detection efficiency.
\\
For single photons entering the interferometer
(Fig. \ref{fig:LLLsystem}) in the absence of an object, the mean photon number at the output ports is shown in Table \ref{Table-input1}.
In the presence of an object, the mean photon number at the output ports is shown in Table  \ref{Table-input2} and the variance in photon number in these tables is $(\Delta^2 N_{k})=
\expval{N_{k}}(1-\expval{N_{k}})$.
\begin{table*}[!htbp]
	\begin{center}
	\caption{
		Mean photon number at the output ports for input states $a_1^\dagger \ket{0}=\ket{1}_{1a}$ and $a_2^\dagger \ket{0}=\ket{1}_{2a}$
		in the absence of an object.
		\label{Table-input1}
	}
		\begin{tabular}{ccc}
			\hline
			Ports \textbackslash \, Measurements
			& Input state $a_1^\dagger \ket{0}=\ket{1}_{1a}$
			& Input state $a_2^\dagger \ket{0}=\ket{1}_{2a}$
			\\ \hline 
			Port $\#3 \,(D_{M_1}) \, \textbackslash \expval{N_{3}}$
			&\small$\tilde{R}\tilde{T}$&$\tilde{T}^2$ \\  
			Port $\#1\, (D_1)\, \textbackslash \expval{N_{1}}$
			&$\tilde{R}^3-2 \tilde{R}^2 \tilde{T} \cos (\varphi )+\tilde{R} \tilde{T}^2$&$4 \tilde{R}^2 \tilde{T} \cos ^2\left(\frac{\varphi }{2}\right)$\\  
			Port $\#2\, (D_2)\, \textbackslash \expval{N_{2}}$
			&$4 \tilde{R}^2 \tilde{T} \cos ^2\left(\frac{\varphi }{2}\right)$ &$\tilde{R}^3-2 \tilde{R}^2 \tilde{T} \cos (\varphi )+\tilde{R}\tilde{T}^2$\\  
			Port $\#4 \,(D_{M_2})\, \textbackslash \expval{N_{4}}$
			&$\tilde{T}^2$&$\tilde{R} \tilde{T}$\\ \hline
		\end{tabular}
\end{center}
\end{table*}	

\begin{table*}[!htbp]
	\begin{center}
	\caption{
		Mean photon number at the output ports for input states $a_1^\dagger \ket{0}=\ket{1}_{1a}$ and $a_2^\dagger \ket{0}=\ket{1}_{2a}$
		in the presence of an object.
		\label{Table-input2}
	}
		\begin{tabular}{ccc}
			\hline
			Ports \textbackslash \, Measurements
			& Input state $a_1^\dagger \ket{0}=\ket{1}_{1a}$
			& Input state $a_2^\dagger \ket{0}=\ket{1}_{2a}$
			\\ \hline 
			Port $\#3 \,(D_{M_1}) \, \textbackslash \expval{N_{3}}$
			&$\tilde{R}\tilde{T}$&$\tilde{T}^2$ \\
			Port $\#1\, (D_1)\, \textbackslash \expval{N_{1}}$
			&$\tilde{R}^3$&$\tilde{R}^2 \tilde{T}$\\
			Port $\#2\, (D_2)\, \textbackslash \expval{N_{2}}$
			&$\tilde{R}^2 \tilde{T}$ &$\tilde{R}\tilde{T}^2$\\
			Port $\#4 \,(D_{M_2})\, \textbackslash \expval{N_{4}}$
			&$\tilde{T}^2$&$\tilde{R} \tilde{T}$\\ \hline
		\end{tabular}
\end{center}
\end{table*}
Note that due to the symmetry of the LLL system, these results are valid for a general beamsplitter, and do not depend on the relative phase between the output modes of the beamsplitter (provided they are identical).
\\\\
We consider two configurations in which a dark port is achievable
and correspondingly, two detection configurations which depend on the transmission ($\tilde{T}$), reflection ($\tilde{R}$) and the phase $\varphi$ of the interferometer.
	In the first configuration we consider, the reflection ($\tilde{R}$) is equal to the transmission ($\tilde{T}$) and is thus referred to as ``symmetric''. In the second configuration we consider $\tilde{R}\neq\tilde{T}$ and is thus referred to as ``asymmetric''.
\\
In addition, since the number of photons at each of the ports 
depend on the physical parameters of the system, we can characterize the system and determine ($\tilde{R},\tilde{T},\cos[2](\frac{\varphi}{2})$) by these measurements (see the Supplement 1).
\subsection{Symmetric}
The ``symmetric'' detection configuration is achieved when 
\begin{equation}\label{key}
	\begin{aligned}
		&\tilde{R}=\tilde{T},&&\varphi = 0.
	\end{aligned}
\end{equation}

These conditions lead to a complete destructive interference at port $\#1$ ($\#2$) for the input state $\ket{1}_{1a}$ ($\ket{1}_{2a}$), which we refer to as a dark port.

This configuration is akin to the well known case considered by Elitzur and Vaidman \cite{elitzur1993quantum}, for a lossless symmetric Mach-Zehnder.

The detection probability $P_{\text{det}}$, namely, the probability to detect an object without absorption by the object is
\begin{equation}\label{key}
	\begin{aligned}
		P_{\text{det}} = \frac{\tau^3}{8}.
	\end{aligned}
\end{equation}
\begin{figure}[h!]
	\begin{center}
		\includegraphics[scale=0.95]{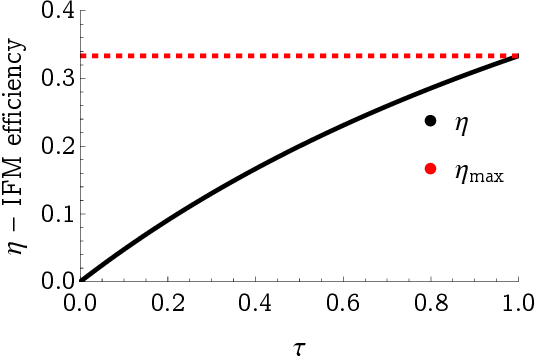}
		\caption{IFM efficiency in the symmetric configuration 
			($\tilde{R}=\tilde{T}=\frac{\tau}{2},\,\varphi=0$) as a function of $\tau$.
			The dashed red line indicates the maximal efficiency $\eta_{\text{max}}=\frac{1}{3}$ in this configuration. 
		}
		\label{IFM-EfficiencyCS}
	\end{center}
\end{figure}
The detection efficiency in this case is  (see Fig. \ref{IFM-EfficiencyCS}) 
\begin{equation}\label{key}
	\begin{aligned}&
		\eta^{(\text{Symmetric})}(\tau)=\frac{\tau}{2+\tau}.
	\end{aligned}
\end{equation}
For the lossless case $\tau\rightarrow1$, we reproduce the known efficiency of $\eta^{(\text{Symmetric})}(\tau\rightarrow1)\equiv\eta_{\text{max}}=\frac{1}{3}$.
It is worth noting that due to symmetry, the efficiency is independent of the input photon port number ($\#1$ or $\#2$).
Furthermore, loss leads to a reduction in the efficiency but not drastically, and one should expect to achieve significant IFM detection in the low loss regime; for $10-15\%$ loss, an efficiency of $\eta \sim 0.29-0.31$ can be reached.
This configuration does not require fine-tuning a phase term and 
is thus experimentally convenient.
\subsection{Asymmetric}
The ``Asymmetric'' detection configuration is achieved 
for $\varphi=\pi$, independently of the values of ($\tilde{R},\tilde{T}$).

\begin{figure}[h!]
	\begin{center}
		\includegraphics[scale=0.95]{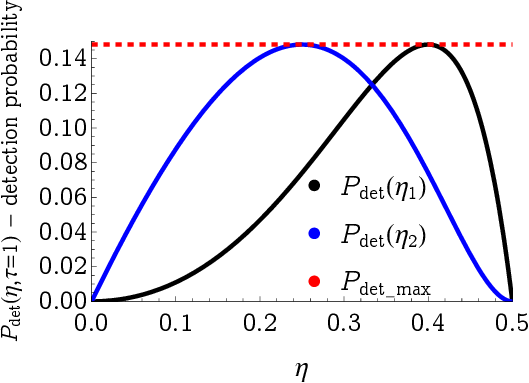}
		\caption{
			\small
			IFM detection probability in the asymmetric configuration as a function of the efficiency $\eta$ in the lossless case ($\tau=1$).
			The black line corresponds to the input state $\ket{1}_{1a}$ and the blue line corresponds to the input state $\ket{1}_{2a}$.
			The dashed red line indicates the maximal detection probability $P_{\text{det}\_\text{max}}=\frac{4}{27}$ in this configuration. 
		}
		\label{IFM-AS-Pd}
	\end{center}
\end{figure}

Following these conditions, port $\#2$ ($\#1$) is dark for the input state $\ket{1}_{1a}$ ($\ket{1}_{2a}$).
In this configuration, the detection efficiency 
given by
\begin{figure}[h!]
	\begin{center}
		\includegraphics[scale=0.95]{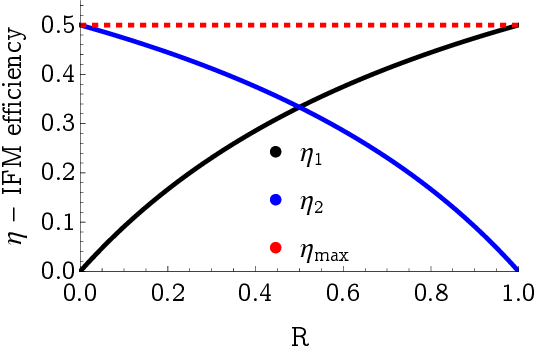}
		\caption{
			\small
			IFM efficiency for the asymmetric detection configuration ($\varphi=\pi$) as a function of $R$ (for $\tau=1$).
			The black line corresponds to the input state $\ket{1}_{1a}$ and the blue line corresponds to the input state $\ket{1}_{2a}$.
			The dashed red line indicates the maximal efficiency $\eta_{\text{max}}=\frac{1}{2}$ in this configuration. 
		}
		\label{IFM-ASEfficiency}
	\end{center}
\end{figure}
\begin{equation}\label{eta-asymmetric}
	\begin{aligned}
		\eta^{(\text{Asymmetric})}(\tilde{R},\tilde{T})=
		\begin{cases}
			\eta_1=\frac{\tilde{R}}{1+\tilde{R}}, & \text{for input }a_1^\dagger\ket{0}=\ket{1}_{1a}
			\\
			\eta_2=\frac{\tilde{T}}{1+\tilde{T}}, & \text{for input }a_2^\dagger\ket{0}=\ket{1}_{2a}
		\end{cases}
	\end{aligned}
\end{equation}
depends on the input state, namely, whether the incident photon entered port $\#1$ ($a_1^\dagger\ket{0}=\ket{1}_{1a}$) or port $\#2$ ($a_2^\dagger\ket{0}=\ket{1}_{2a}$), and on the values of $(\tilde{R},\tilde{T})$
as illustrated in Fig. \ref{IFM-ASEfficiency} for the lossless case.
In the asymmetric dark port setting, it is possible to reach an efficiency of $\eta_{\text{max}}=\frac{1}{2}$ in the lossless case when $\eta_1(\tilde{R}\rightarrow 1)$ or $\eta_2(\tilde{T}\rightarrow 1)$.
While the asymmetric dark port setting is preferable in terms of IFM efficiency, it might be more challenging for experimental realization, since it requires an exact phase term without introducing additional loss.
Conversely, since there is no restriction on the ratio $\tilde{R}/	\tilde{T}$, it is preferable in experimental scenarios where the control of this ratio is limited. 
Furthermore, high efficiencies can be reached in the presence of low-loss; 
for $10-15\%$ loss, an efficiency of $\eta \sim 0.45-0.47$ can be reached, provided that $\tilde{R}/	\tilde{T}$ is either very small or very large.
\\\\
While the maximal efficiency can be reached in either choice of input, the relation between the detection probability $P_{\text{det}}$ and the efficiency $\eta$, does depend on the choice of input, as shown in Fig. \ref{IFM-AS-Pd}.
The detection probability $P_{\text{det}}$ 
can be expressed by 
the efficiency 
and loss using Eq. (\ref{RT-coeff}, \ref{eta-asymmetric}) and it
is given by
\begin{equation}\label{key}
	\begin{aligned}
		P_{\text{det}}=\tilde{R}^2 \tilde{T}
		&=\left(\frac{\eta_1}{1-\eta_1}\right)^{2}\left(\tau-\frac{\eta_1}{1-\eta_1}\right)
		\\&=
		\left(\tau-\frac{\eta_2}{1-\eta_2}\right)^{2}\frac{\eta_2}{1-\eta_2}	.
	\end{aligned}
\end{equation}
In a typical experiment of this sort, there is a fixed average flux of heralded single photons determining the minimal experiment runtime, which is also dictated by the number of measurements required for achieving the desired SNR. Since the detection probability is related to the experiment runtime, it is preferable to work in a regime where both the detection probability and the efficiency are high.
The optimal scenario is therefore to work in a regime where $\tilde{R}>\tilde{T}$ with the input state $\ket{1}_{1a}$ as can be seen in Fig. \ref{IFM-AS-Pd} for this case ($P_{\text{det}}(\eta_1)$)
.
If however, due to technical experimental limitations, it is only possible to work in a regime where $\tilde{R}<\tilde{T}$, then the desired input state is $\ket{1}_{2a}$ to ensure an efficiency which is higher than the efficiency in the symmetric case
$
\frac{\tau}{2+\tau}
$
. 
Note that in the particular case where $\tilde{R}=\tilde{T} \Leftrightarrow\eta_1=\eta_2$ we revert to the symmetric detection configuration with 
detection probability of $\frac{\tau^3}{8}$ and
switched ports due to the $\pi$ shift.
Furthermore, since the detection probability diminishes rapidly as one attempts to reach the maximal efficiency, it is suggested to work in a regime where the detection probability and the efficiency are both high.
In fact, there is a ``sweet spot'' in which both the detection probability and the efficiency surpass what can be achieved in the symmetric configuration.
\begin{equation}\label{key}
	\begin{aligned}
		1<&\frac{\tilde{R}}{\tilde{T}}<2+\sqrt{5},
		\\
		\frac{\tau}{2+\tau}<&\eta_1<\frac{(2+\sqrt{5})\tau}{3+\sqrt{5}+(2+\sqrt{5})\tau}.
	\end{aligned}
\end{equation}
This region starts at the intersection point of $\eta_1=\eta_2 \Rightarrow P_{\text{det}}=\frac{\tau^3}{8}$ and ends where 
the detection probability of $\eta_1$ returns back to the same value.
To illustrate this point in the lossless case, it is possible to achieve efficiency of $\eta_1 \approx 0.44$ with detection probability of $P_{\text{det}}=\frac{1}{8}$ 
by tuning $R\rightarrow\frac{1+\sqrt{5}}{4}$.

\section{Lossy beamsplitter modeling based on Laue diffraction}
\label{section:Laue-beamsplitter}
\begin{figure*}
	\includegraphics[scale=0.9]{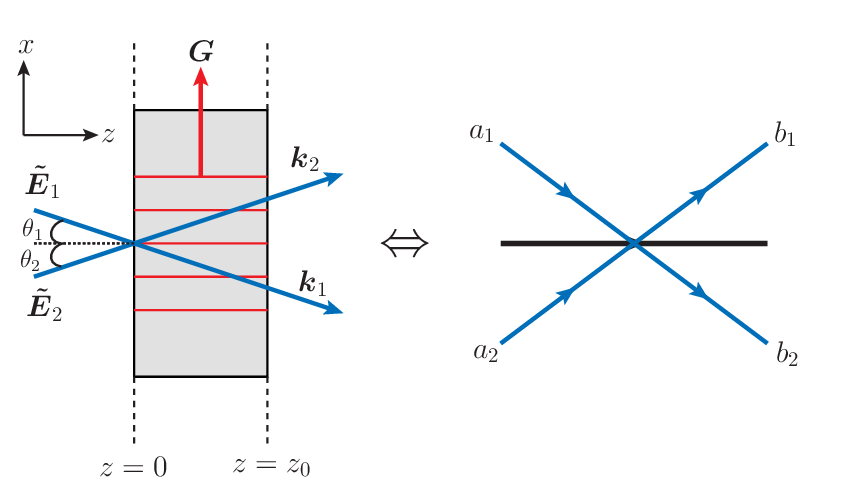}
	\caption{
		\small
		On the left, fields
		$(\vb*{\tilde{E}}_1,\, \vb*{\tilde{E}}_2)$ with wave vectors $(\vb*{k}_1,\, \vb*{k}_2)$ at incident angles $(\theta_1,\, \theta_2)$ respectively, undergo Laue diffraction with respect to the reciprocal lattice vector $\vb*{G}$ by a crystal of width $z_0$.
		On the right, analogous representation of Laue diffraction as a beamsplitter in which input operators $(a_1,\, a_2)$ are transformed to output operators $(b_1,\, b_2)$.
	}
	\label{fig:Laue-beamsplitter}	
\end{figure*}
\begin{figure}[h!]
	\begin{center}
		\includegraphics[scale=0.93]{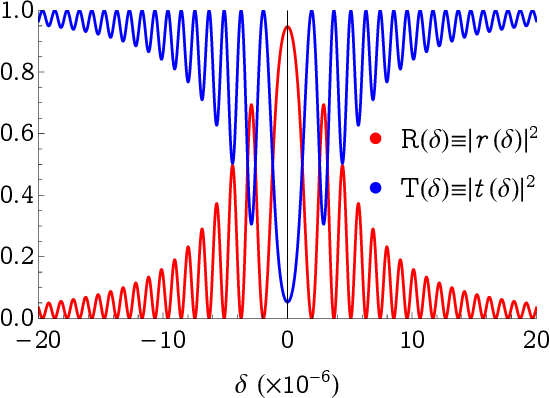}
		\caption{
			\small
			Angular dependence of the lossless reflection ($R$) and transmission ($T$) coefficients (see Eq. (\ref{RT-coeff})) on the mismatch angle $\delta$ in units of $\mu \text{rad}$ for silicon crystal in the low-loss regime
			($\tau \approx 0.994$).
		 	}
		\label{RT-delta-dependence}
	\end{center}
\end{figure}

The theoretical description shown in previous sections
relied on the possible modeling of Laue diffraction as a transfer matrix between input and output field modes (see Fig. \ref{fig:Laue-beamsplitter}).
To accommodate this general formalism for x-ray optics, here we show such modeling.
It is based on the quantized Takagi-Taupin equations \cite{authier2004dynamical} for the field operators $a_1(z)$ and $a_2(z)$ accompanied by Langevin operators to account for loss.
\begin{equation}\label{Lau-equations}
	\begin{aligned}&
		\pdv{a_1(z)}{z}=-\left(\alpha+\frac{i\Delta k_z}{2}\right)a_1(z)+i\kappa a_2(z)+\sqrt{2\alpha}f_1(z),
		\\&
		\pdv{a_2(z)}{z}=-\left(\alpha-\frac{i\Delta k_z}{2}\right)a_2(z)+i\kappa a_1(z)+\sqrt{2\alpha}f_2(z).
	\end{aligned}
\end{equation}
The coefficients $\alpha,\,\kappa,\,\Delta k_z$ are the absorption coefficient, coupling coefficient and  phase mismatch along the optical axis respectively.
Their explicit form is given by
\begin{equation}\label{key}
	\begin{aligned}
		\alpha&=\tilde{\sigma}_0\omega \gamma \rho_0,
		\\
		\kappa&=\tilde{\sigma}_0(\omega^2-\omega_0^2)\rho_{\vb*{G}},
		\\
		\Delta k_z&=(\vb*{k}_2-\vb*{k}_1-\vb*{G})\cdot \hat{z},
		\\
		\tilde{\sigma}_0 &=-
		\frac{e \mu_0 c}{2m_e n \cos(\theta_b)}
		\frac{\omega}{\omega^2(1+\gamma^2)-\omega_0^2}.
	\end{aligned}
\end{equation}
When the incident angles slightly deviate from Bragg's condition by a small parameter $\delta$, such that $\theta_1=\theta_b+\delta ,\,\theta_2=\theta_b-\delta$, the phase mismatch can be approximated $\Delta k_z \approx\abs{\vb*{G}}\delta $.
The Lorentz model was used to describe the linear response for a periodic charge density
\begin{equation}\label{key}
	\rho(\vb*{x})=\sum_{\vb*{G}}\rho_{\vb*{G}}e^{i\vb*{G}\cdot \vb*{x}},
\end{equation} 
such that $\rho_{\vb*{G}}$ is the charge density Fourier component associated with reciprocal lattice vector $\vb*{G}$ and $\rho_{0}$ is the mean electron charge density.
The parameters $-e,\, m_e, \,c ,\,\mu_0,\,\omega_0,\, \omega,\,\gamma,n,\, \theta_b$ are the electron's charge, electron's mass, speed of light in vacuum, permeability of free space, resonance frequency of the Lorentz oscillator, field's frequency, damping coefficient, refraction index and Bragg's angle, respectively. 
Following the solution of these equations, the relation between the input operators $\vb*{a}=(a_1(0)\equiv a_1,\,a_2(0)\equiv a_2)$ and the output operators $(a_1(z_0),\,a_2(z_0))$ at $z_0$ is given by
\begin{equation}\label{key}
	\begin{aligned}
		a_1(z_0)&=e^{-\alpha z_0}
		\left(
		t^*(z_0)a_1 +i r(z_0)a_2
		\right)
		+
		\sqrt{1-e^{-2\alpha z_0}}a_{n1},
		\\
		a_2(z_0)&=e^{-\alpha z_0}
		\left(
		i r(z_0)a_1 +t(z_0)a_2
		\right)
		+
		\sqrt{1-e^{-2\alpha z_0}}a_{n2},
	\end{aligned}
\end{equation}
where
\begin{equation}\label{key}
	\begin{aligned}
		&t(z)=\sech(\phi)\cos(\cosh(\phi)\kappa z+i\phi),
		\\
		&r(z)=\sech(\phi)\sin(\cosh(\phi)\kappa z),
	\end{aligned}
\end{equation}
and $\sinh(\phi)=-\frac{\Delta k_z}{2\kappa}$.
The addition of the bosonic noise operators $\vb*{a}_n=(a_{n1},a_{n2})$ is required to ensure the commutation relations and their contribution is proportional to the loss.
Rewriting this relation in terms of the output operators $\vb*{b}=(b_1, b_2)$
as depicted in Fig. \ref{fig:Laue-beamsplitter} (on the right) yields
\begin{equation}\label{key}
	\vb*{b}(z)=e^{-\alpha z} \text{bs}(z)\vb*{a}+\sqrt{1-e^{-2\alpha z}}\vb*{a}_n,
\end{equation}
where
\begin{equation}\label{key}
	\text{bs}(z)=\mqty(i r(z) & t(z)
	\\
	t^*(z) & i r(z))	.
\end{equation}
The transformation consists of two parts.
The first term corresponds to a standard beamsplitter transformation ($\text{bs}(z)$) which is attenuated by a factor of $e^{-\alpha z}$ due to absorption.
The presence of loss gives rise to the second term which describes the additional noise.
Controlling the transmission $T=\abs{t}^2$ and reflection $R=\abs{r}^2$ coefficients can be achieved by tuning the incident angle and thus modifying $\Delta k_z$.
To illustrate this point, we show the angular dependence of ($R,T$) on the mismatch angle $\delta$
in Fig. \ref{RT-delta-dependence}, in the case of silicon with $z_0=200\,\mu m$, photon energy of $\hbar \omega = 30\,\text{keV}$ and diffraction plane of (2, 2, 0).
In this example, $e^{-2\alpha z_0} \approx 0.994$ therefore, the loss is negligible and adequate control on $R,T$ is possible by angular deviations on the order of $\sim \mu \, \text{rad}$.
\begin{figure}[h!]
	\begin{center}
		\includegraphics[scale=0.93]{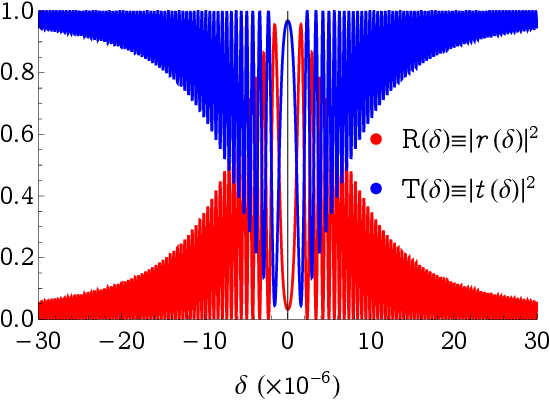}
		\caption{
			\small
			Angular dependence of the lossless reflection ($R$) and transmission $(T)$ coefficients (see Eq. (\ref{RT-coeff})) on the mismatch angle $\delta$ in units of $\mu \text{rad}$ for silicon crystal in the lossy regime ($\tau \approx 0.532$). 
     		}
		\label{RT-delta-dependence2}
	\end{center}
\end{figure}

Furthermore, adequate tunability of $R,T$ is feasible in the lossy domain as depicted in Fig. \ref{RT-delta-dependence2}, which corresponds to the case of silicon with $z_0=500\,\mu m$, photon energy of $\hbar \omega = 18\,\text{keV}$, diffraction plane of (2, 2, 0) and absorption of $1-e^{-2\alpha z_0}\sim 0.468$. 
In this case, the oscillations frequency is larger which provides finer tunability of $R,T$.
Despite the substantial loss ($\tau \approx 0.532$), our approach demonstrates notable resilience with achievable IFM efficiencies of approximately $\eta^{(\text{Symmetric})}\sim 0.21$ and $\eta^{(\text{Asymmetric})}\sim 0.34$.
This example shows that high IFM efficiency can be obtained even when there is substantial loss.
\section{Discussion}
We have shown that x-ray IFM with the LLL interferometer is feasible even with significant loss.
Our analysis highlights the robustness of IFM to loss.
We have analyzed two IFM detection configurations with efficiencies ranging from $\frac{1}{3}$ to $\frac{1}{2}$.

The symmetric configuration is slightly simpler to implement, since it does not require fine-tuning phase.
In this configuration, best-case efficiency can asymptotically reach  $\eta \rightarrow\frac{1}{3}$ (in the absence of loss). In the presence of low-loss ($10-15\%$), reasonable efficiencies can still be reached ($\eta \sim 0.29-0.31$) and thus it is a promising experimental configuration, provided there is an adequate control on the reflection ($\tilde{R}$) and transmission ($\tilde{T}$) coefficients.
The asymmetric configuration is preferable for a scenario in which a well controlled phase
can be introduced (without adding loss) to the setup.
Additionally, it can be important for cases in which there is limited control of the ratio between the reflection and transmission coefficients ($\tilde{R},\,\tilde{T}$).
In this configuration, best-case efficiency can reach asymptotically $\eta \rightarrow\frac{1}{2}$ (in the absence of loss), if the ratio $\frac{\tilde{R}}{\tilde{T}}$ can be controlled.
In the presence of low-loss ($10-15\%$), good efficiencies can be reached ($\eta \sim 0.45-0.47$) and thus can be used to demonstrate efficient x-ray IFM in the presence of loss.

Furthermore, due to the symmetry of the LLL setup, the results presented here are independent of the way one models the beamsplitters (provided they are identical).
However, in order to properly design an LLL system in practice,		
it is instructive to have a model of a lossy beamsplitter which is based on Laue diffraction. For this purpose, we constructed such a model and showed that adequate control of the ratio $\tilde{R}/\tilde{T}$ is possible while remaining in the low loss regime as well as in the lossy regime. 
Such tunability suggests that x-ray IFM can be demonstrated with high efficiency in the presence of significant loss in the system.

Additionally, we have presented a convenient method to experimentally characterize the properties of the interferometer. This method allow for the determination of the key parameters $\tilde{R},\tilde{T}$ and $\cos[2](\frac{\varphi}{2})$, using only the measurements of the outgoing ports.

In conclusion,
our study demonstrates the feasibility of x-ray IFM using the LLL interferometer.   
A successful demonstration of IFM in the x-ray regime has great potential for delving into more advanced schemes for high efficiency IFM, e.g. high efficiency interrogation and imaging based on the quantum Zeno effect, as well as x-ray IFM for semi-transparent objects. 
These advancements have the potential to revolutionize the development of low-dose x-ray imaging techniques based on IFM.
\bibliography{XR-IFM}
\end{document}